\documentclass[runningheads]{llncs}
\usepackage{graphicx}
\usepackage{multirow}
\usepackage{amssymb}
\usepackage{booktabs}
\usepackage{amsmath}
\usepackage{bbm}
\usepackage{threeparttable} % write table foot
\begin{document}
\title{Deep-OCTA: Ensemble Deep Learning Approaches for Diabetic Retinopathy Analysis on OCTA Images
% \thanks{Supported by organization x.}
}
\titlerunning{Deep-OCTA}
% If the paper title is too long for the running head, you can set
% an abbreviated paper title here
%
% \author{First Author\inst{1}\orcidID{0000-1111-2222-3333} \and
% Second Author\inst{2,3}\orcidID{1111-2222-3333-4444} \and
% Third Author\inst{3}\orcidID{2222--3333-4444-5555}}
\author{Junlin Hou\inst{1} \and
Fan Xiao\inst{2} \and
Jilan Xu\inst{1} \and Yuejie Zhang\inst{1}* \and \\ Haidong Zou\inst{3} \and Rui Feng\inst{1,2}*}
\authorrunning{J. Hou et al.}
% First names are abbreviated in the running head.
% If there are more than two authors, 'et al.' is used.
%
\institute{School of Computer Science, Shanghai Key Laboratory of Intelligent Information Processing, Fudan University, China\\ \email{\{jlhou18,jilanxu18,yjzhang,fengrui\}@fudan.edu.cn}\\
\and
Academy for Engineering and Technology, Fudan University, China\\
\email{21210860085@m.fudan.edu.cn}\\
\and
Department of Ophthalmology, Shanghai General Hospital, School of Medicine, Shanghai Jiao Tong University, China\\
\email{zouhaidong@sjtu.edu.cn}}

% \institute{Princeton University, Princeton NJ 08544, USA \and
% Springer Heidelberg, Tiergartenstr. 17, 69121 Heidelberg, Germany
% \email{lncs@springer.com}\\
% \url{http://www.springer.com/gp/computer-science/lncs} \and
% ABC Institute, Rupert-Karls-University Heidelberg, Heidelberg, Germany\\
% \email{\{abc,lncs\}@uni-heidelberg.de}}
%
\maketitle              % typeset the header of the contribution
\begin{abstract}
The ultra-wide optical coherence tomography angiography (OCTA) has become an important imaging modality in diabetic retinopathy (DR) diagnosis. However, there are few researches focusing on automatic DR analysis using ultra-wide OCTA. In this paper, we present novel and practical deep-learning solutions based on ultra-wide OCTA for the Diabetic Retinopathy Analysis Challenge (DRAC). In the segmentation of DR lesions task, we utilize UNet and UNet++ to segment three lesions with strong data augmentation and model ensemble. In the image quality assessment task, we create an ensemble of Inception-V3, SE-ResNeXt, and Vision Transformer models. Pre-training on the large dataset as well as the hybrid MixUp and CutMix strategy are both adopted to boost the generalization ability of our model. In the DR grading task, we build a Vision Transformer (ViT) and find that the ViT model pre-trained on color fundus images serves as a useful substrate for OCTA images. Our proposed methods ranked 4th, 3rd, and 5th on the three leaderboards of DRAC, respectively. 
The source code will be made available at \url{https://github.com/FDU-VTS/DRAC}.

\end{abstract}

\section{Introduction}
Diabetic Retinopathy (DR) is a chronic progressive disease that causes visual impairment due to retinal microvascular damage. It has become a leading cause of legal blindness in the working-age population worldwide \cite{zheng2012worldwide}. DR is diagnosed by the presence of retinal lesions, such as microaneurysms (MAs), intraretinal microvascular abnormalities (IRMAs), nonperfusion areas (NPAs) and neovascularization (NV) \cite{bin_sheng_2022_6362349}. The traditional diagnosis of DR grading mainly relies on fundus photography and fluorescein fundus angiography (FFA). With rising popularity, OCT angiography (OCTA) has become a reliable tool due to its capability of visualizing the retinal and choroidal vasculature at a microvascular level \cite{spaide2018optical}. Compared to fundus photography and FFA, the ultra-wide OCTA can non-invasively detect the changes of DR neovascularization, thus it is an important imaging modality to assist ophthalmologists in diagnosing Proliferative DR (PDR). 

Recently, deep learning approaches have achieved promising performance in DR diagnosis. Based on fundus photography, numerous deep learning methods are proposed for lesion segmentation, biomarkers segmentation, disease diagnosis and image synthesis \cite{li2021applications}. A deep learning system, named DeepDR, is developed to perform real-time image quality assessment, lesion detection and DR grading \cite{dai2021deep}. However, there are currently few works capable of automatic DR analysis using ultra-wide OCTA images.

\begin{figure}[t]
\centering
\includegraphics[width=0.9\textwidth]{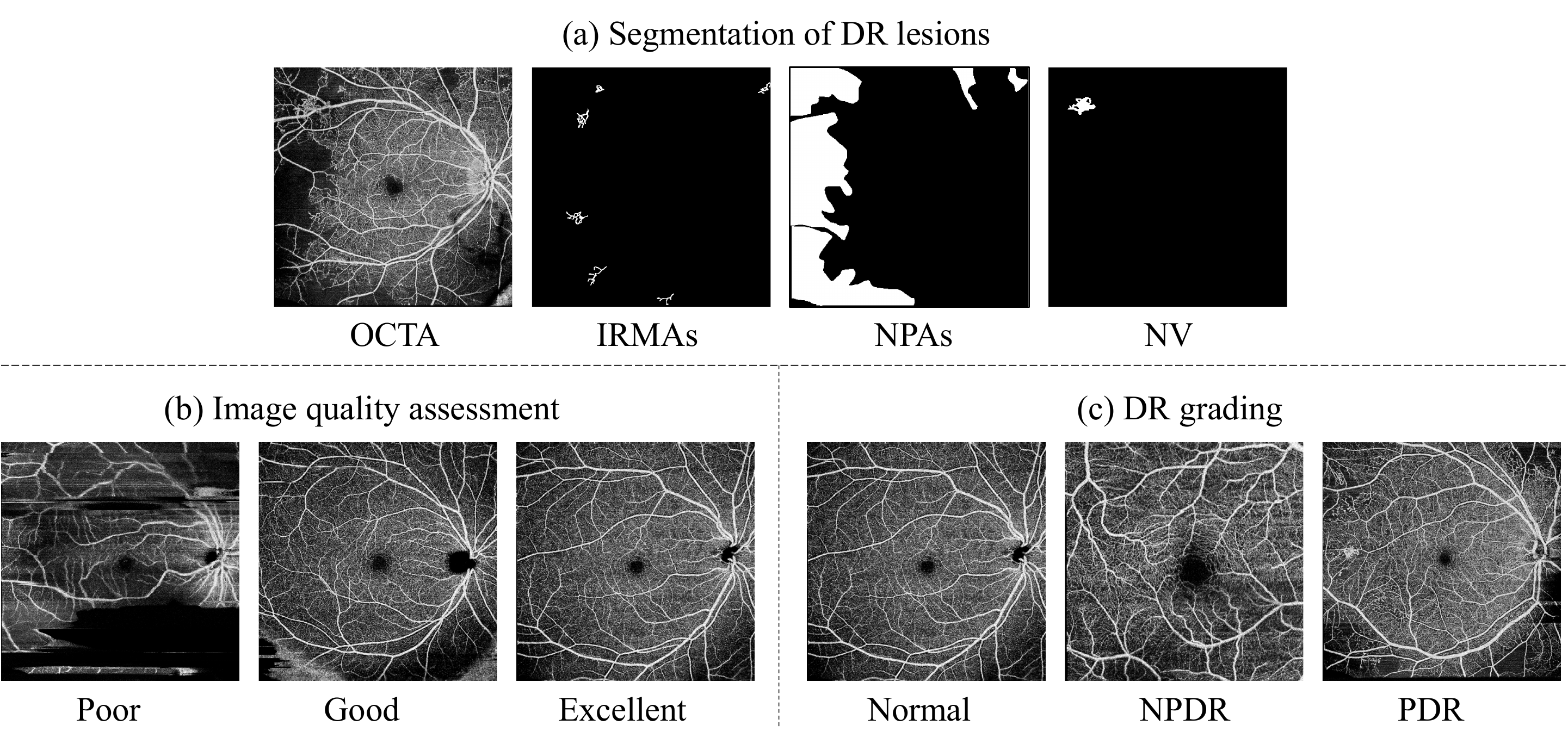}
\caption{Samples from three tasks in the DRAC dataset, including (a) segmentation of DR lesions, (b) image quality assessment, and (c) DR grading.} 
\label{intro}
\end{figure}

In this paper, we propose novel and practical methods for DR analysis based on ultra-wide OCTA images. Our methods are demonstrated effective in the MICCAI 2022 Diabetic Retinopathy Analysis Challenge (DRAC) \cite{bin_sheng_2022_6362349}. This challenge provides a standardized ultra-wide OCTA dataset, including lesion segmentation, image quality assessment and DR grading. As illustrated in Fig. \ref{intro}, the segmentation subset includes three DR lesions, i.e. Intraretinal Microvascular Abnormalities (IRMAs), Nonperfusion Areas (NPAs), and Neovascularization (NV). The image quality assessment subset 
divides images into three quality levels, i.e. Poor quality level, Good quality level, and Excellent quality level. The DR grading subset contains images of three different DR grades, i.e. Normal, Non-Proliferative DR (NPDR), and Proliferative DR (PDR). 
In our approach, we first develop two encoder-decoder networks, namely UNet \cite{ronneberger2015u} and UNet++ \cite{zhou2018unet++}, to segment three lesions separately. Strong data augmentation and model ensemble are demonstrated helpful to the generalization ability of our models. For image quality assessment, we create an ensemble of multiple state-of-the-art neural networks. The networks are first pre-trained on large-scale datasets, and then fine-tuned on DRAC dataset using the hybrid MixUp and CutMix strategy. For DR grading, we adopt a Vision Transformer model, which utilizes self-attention to integrate information across the entire OCTA image. Extensive experiments on DRAC dataset show that our proposed solutions achieve superior DR analysis performance, ranking 4th, 3rd, 5th on the three leaderboards, respectively.

\section{Segmentation of Diabetic Retinopathy Lesions}

\subsection{Methodology}
We adopt UNet \cite{ronneberger2015u} and UNet++ \cite{zhou2018unet++} networks with pre-trained encoders for DR lesion segmentation. 
Customized strategies are designed to train the segmentation models of three different lesions.
For IRMAs segmentation, Step1 learning rate schedule and color jittering augmentation are both adopted. However, they are not utilized to boost the segmentation performance of NPAs and NV lesions. Besides, we employ the model ensemble strategy when predicting the segmented masks of IRMAs and NPAs. In the following section, we will introduce the UNet and UNet++ networks, learning rate schedules, and loss functions.

\subsubsection{Backbone network.} 
(1) UNet \cite{ronneberger2015u} is an encoder-decoder network, where the encoder includes down-sampling layers and the decoder consists of up-sampling layers with skip connections. The feature structures of different levels are combined through skip connections.
(2) UNet++ \cite{zhou2018unet++} proposes an efficient ensemble of U-Nets of varying depths, which partially share an encoder and co-learn simultaneously using deep supervision. It redesigns skip connections to exploit multi-scale semantic features, leading to a flexible feature aggregation scheme. 

% based on UNet utilizes long and short skip connection in order to reduce the scale difference of feature map in fusion, meanwhile, leads in more parameters. It is no longer a fixed depth structure with UNet embedded at different depths. Simultaneous training of embedded UNets with different depths leads to co-training among UNets, resulting in better performance.

\subsubsection{Learning rate schedule.}
In order to train our network efficiently, we analyze two different schedules for learning rate decay. One method, named Step1, divides the initial learning rate by 10 at 25\% of the total number of training epochs. The other method, named Step2, is to decay the learning rate to 0.6$\times$ of the previous value every quarter epoch of the total epochs.

\subsubsection{Loss function.}
We adopt two loss functions, i.e. Dice loss $\mathcal{L}_{D}$  and Jaccard loss $\mathcal{L}_{J}$, to train the segmentation models of DR lesions. Specifically, the Dice loss $\mathcal{L}_{D}$ is expressed by:
\begin{equation}
    \mathcal{L}_{D}=\frac{1}{N}\sum_{i=1}^N \left(1-\mathrm{Score}_{Dice(i)}\right),
\end{equation}
where $N$ denotes the total number of samples, $\mathrm{Score}_{Dice(i)}$ is the Dice score of the sample $x_i$.
The Jaccard loss $\mathcal{L}_{J}$ is defined as:
\begin{equation}
    \mathcal{L}_{J}=\frac{1}{N}\sum_{i=1}^N \left(1-\mathrm{Score}_{IoU(i)}\right),
\end{equation}
where $\mathrm{Score}_{IoU(i)}$ denotes the IoU score of the sample $x_i$.
We train the networks by each loss function and joint loss function for segmentation of three lesions.

\subsection{Dataset}
\subsubsection{DRAC dataset}
The DRAC dataset for segmentation of diabetic retinopathy lesions contains three different diabetic retinopathy lesions, i.e. Intraretinal Microvascular Abnormalities (1), Nonperfusion Areas (2), Neovascularization (3). Training set consists of 109 images and corresponding lesion masks, where each category includes 86/106/35 images, respectively. We randomly select 20\% of the training set as the validation set. Testing set consists of 65 images and the ground-truth masks are not available during the challenge.

\subsection{Implementation details}
All images with the original image size (1024$\times$1024) are fed to UNet and UNet++. Strong data augmentation includes horizontal flipping, rotating, random cropping, gaussian noise, perspective, and color jittering. The networks are optimized using the Adam algorithm and trained for 100 epochs. The initial learning rate is set to 1e-4, and we use step learning rate schedule.

\subsection{Experimental results}

\setlength{\tabcolsep}{4pt}
\begin{table}[t]
\begin{center}
\caption{The results on the DRAC dataset of IRMAs segmentation.}
\label{table:result task2 of IRMA}
\resizebox{\linewidth}{!}{
\begin{tabular}{c|l|c|c|c|c|c|c}
\hline
ID & Method & Loss & Metric & Schedule & Aug & Val Dice & Test Dice\\
\hline
1 & UNet++ & Dice & Dice & Step1 & All & 0.5637 & 0.4172\\
2 & UNet++ & Dice & IoU & Step1 & All & 0.6145 & 0.4166\\
3 & UNet++ & Jaccard & IoU & Step1 & All & 0.4648 & 0.3933\\
4 & UNet++ & Dice & Dice & Step2 & All & 0.6139 & 0.3788\\
5 & UNet & Dice & Dice & Step2 & All & 0.5215 & 0.3514\\
6 & UNet & Dice & Dice & Step2 & No CJ & 0.6115 & 0.2774\\
% 4 & UNet++ & Dice & Dice & Step2 & All & 0.5446 & 0.3509\\
% 5 & UNet++ & Dice & Dice & Step2 & Weak CJ & 0.6139 & 0.3788\\
% 6 & UNet & Dice & Dice & Step2 & Weak CJ & 0.5215 & 0.3514\\
% 7 & UNet & Dice & Dice & Step2 & No CJ & 0.6115 & 0.2774\\
% \multirow{2}{*}{8} & Ensemble & \multirow{2}{*}{-} & \multirow{2}{*}{-} & \multirow{2}{*}{-} & \multirow{2}{*}{-} & \multirow{2}{*}{-} & \multirow{2}{*}{\textbf{0.4257}}\\
%  & ID `1,2,5' & & & & & &\\
7 & Ensemble ID `1,2,4' & - & - & - & - & - & \textbf{0.4257}\\
\hline
\end{tabular}
}
\end{center}
\end{table}
\setlength{\tabcolsep}{1.4pt}

\setlength{\tabcolsep}{4pt}
\begin{table}[t]
\begin{center}
\caption{The results on the DRAC dataset of NPAs segmentation. }
\label{table:result task2 of NPA}
\resizebox{\linewidth}{!}{
\begin{threeparttable}
\begin{tabular}{c|l|c|c|c|c|c|c}
\hline
ID & Method & Loss & Metrics & Schedule & Aug & Val Dice & Test Dice\\
\hline
1 & UNet++ & Dice & Dice & Step1 & All & 0.6211 & 0.5874\\
2 & UNet++ & Dice & IoU & Step1 & All & 0.6460 & 0.5942\\
3 & UNet++ & Jaccard & IoU & Step1 & All & 0.6348 & 0.5930\\
4 & UNet++ & Dice & Dice & Step2 & All & 0.6699 & 0.6105\\
5 & UNet & Dice & Dice & Step2 & All & 0.6805 & 0.6197\\
6\tnote{$\dag$} & UNet & Dice & Dice & Step2 & All & 0.7027 & 0.6275\\
% 5 & UNet++ & Dice & Dice & Step2 & Weak CJ & 0.6635 & 0.6015\\
% 6 & UNet & Dice & Dice & Step2 & Weak CJ & 0.6805 & 0.6197\\
% 7\tnote{1} & UNet & Dice & Dice & Step2 & Weak CJ & 0.7027 & 0.6275\\
% \multirow{2}{*}{8} & \multirow{2}{*}{UNet} & Dice+ & \multirow{2}{*}{dice} & \multirow{2}{*}{Step} & \multirow{2}{*}{Weak CJ} & \multirow{2}{*}{0.6541} & \multirow{2}{*}{0.6011}\\
%  & & Jaccard & & & & &\\
% 8 & UNet & Dice\&Jaccard & Dice & Step2 & Weak CJ & 0.6541 & 0.6011\\
7 & UNet & Dice\&Jaccard & Dice & Step2 & All & 0.6541 & 0.6011\\
8 & UNet & Dice & Dice & Step2 & No CJ & 0.6434 & 0.6261\\

% \multirow{2}{*}{{10}} & Ensemble & \multirow{2}{*}{-} & \multirow{2}{*}{-} & \multirow{2}{*}{-} & \multirow{2}{*}{-} & \multirow{2}{*}{-} & \multirow{2}{*}{\textbf{0.6414}}\\
%  & ID `4,6,7,8 & & & & & & \\
9 & Ensemble ID `4,5,6,7' & - & - & - & - & - & \textbf{0.6414}\\
\hline
\end{tabular}
\begin{tablenotes}
    \item{$\dag$} The network was re-trained by a different random seed.
\end{tablenotes}
\end{threeparttable}

}
\end{center}

\end{table}
\setlength{\tabcolsep}{1.4pt}

\setlength{\tabcolsep}{4pt}
\begin{table}[t]
\begin{center}
\caption{The results on the DRAC dataset of NV segmentation. }
\label{table:result task2 of NV}
% \resizebox{\linewidth}{!}{
\begin{tabular}{c|l|c|c|c|c|c|c}
\hline
ID & Method & Loss & Metrics & Schedule & Aug & Val Dice & Test Dice\\
\hline
1 & UNet++ & Dice & Dice & Step1 & All & 0.2360 & 0.3838\\
2 & UNet++ & Jaccard & IoU & Step1 & All & 0.1984 & 0.3741\\
3 & UNet++ & Dice & Dice & Step2 & All & 0.5250 & 0.4803\\
4 & UNet & Dice & Dice & Step2 & All & 0.5547 & 0.5445\\
% 4 & UNet++ & Dice & Dice & Step2 & Weak CJ & 0.3591 & 0.4607\\
% 5 & UNet & Dice & Dice & Step2 & Weak CJ & 0.5547 & 0.5445\\
5 & UNet & Dice & Dice & Step2 & No CJ & 0.4881 & \textbf{0.5803}\\
\hline
\end{tabular}
% }
\end{center}
\end{table}
\setlength{\tabcolsep}{1.4pt}

\setlength{\tabcolsep}{4pt}
\begin{table}[t]
\begin{center}
\caption{The top-5 results on the leaderboard of segmentation of DR lesions.}
\label{table:leaderboard task1}
\resizebox{\linewidth}{!}{
\begin{tabular}{llccccccccc}
\hline\noalign{\smallskip}
\multirow{2}{*}{Rank} & \multirow{2}{*}{Team} & \multirow{2}{*}{mDice} & \multirow{2}{*}{mIoU} & \multicolumn{2}{c}{IRMAs} & \multicolumn{2}{c}{NPAs} & \multicolumn{2}{c}{NV} \\
\cmidrule(r){5-6} \cmidrule(r){7-8} \cmidrule(r){9-10}
 & & & & Dice & IoU & Dice & IoU & Dice & IoU \\
\noalign{\smallskip}
\hline
1 & FAI & 0.6067 & 0.4590 & 0.4704 & 0.3189 & 0.6926 & 0.5566 & 0.6571 & 0.5015\\
2 & KT\_Bio\_Health & 0.6046 & 0.4573 & 0.4832 & 0.3299 & 0.6736 & 0.5396 & 0.6570 & 0.5024\\ 
3 & AiLs & 0.5756 & 0.4223 & 0.4672 & 0.3144 & 0.6680 & 0.5260 & 0.5917 & 0.4433\\
4 & FDVTS\_DR (ours) & 0.5491 & 0.4041 & 0.4257 & 0.2853 & 0.6414 & 0.5031 & 0.5803 & 0.4240\\
5 & LaTIM & 0.5387 & 0.3966 & 0.4079 & 0.2691 & 0.6515 & 0.5129 & 0.5566 & 0.4077\\
\hline
\end{tabular}
}
\end{center}
\end{table}
\setlength{\tabcolsep}{1.4pt}

\subsubsection{Backbone network.}
Both UNet and UNet++ backbone structures are used to segment three DR lesions separately. From the results in the 4th and 5th rows of Table \ref{table:result task2 of IRMA}, we can see that UNet++ obtains superior IRMAs segmentation results than UNet by 2.74\% test dice. On the contrary, UNet is demonstrated more suitable for the NPAs and NV segmentation. It outperforms UNet++ by a 1.70\% improvement for NPAs and a 6.42\% improvement for NV, as can be seen from the 4th and 6th rows of Table \ref{table:result task2 of NPA} and the 3rd and 4th rows of Table \ref{table:result task2 of NV}.

\subsubsection{Learning rate schedule.}
% While the Step1 schedule helps to improve 0.0663 test Dice in the 4th and 1st rows of Table \ref{table:result task2 of IRMA}, the Step2 schedule contributes to increasing 0.0231 and 0.0965 test Dice performance respectively in the 1st and 4th rows of Table \ref{table:result task2 of NPA} and the 1st rows and 3rd rows of Table \ref{table:result task2 of NV}. Thus, we can find that Step1 schedule is useful for training of IRMAs and the Step2 schedule is better than Step1 on training of two other lesions.

We compare the impact of different learning rate schedules for each lesion segmentation. As can be seen in the 1st and 4th rows in Table \ref{table:result task2 of IRMA}, the Step1 schedule largely outperforms Step2 by 3.84\% test dice for IRMAs segmentation. In contrast, the Step1 schedule fails in segmenting the other two lesions. From the 1st and 4th rows of Table \ref{table:result task2 of NPA}, we can see that the NPAs segmentation performance obtained by the Step2 schedule greatly surpasses Step1 by 2.31\% test dice. For NV segmentation, the results in the 1st and 3rd rows of Table \ref{table:result task2 of NV} show that the Step2 schedule obtains a significant improvement of 9.65\% on test dice.

\subsubsection{Color jittering.}
% It can be seen from Table \ref{table:result task2 of IRMA} that Color Jittering (CJ) can effectively solve the overfitting problem of segmentation of IRMAs. Comparing with the 5th and 6th rows of Table \ref{table:result task2 of IRMA}, we can see the test dice increase 0.074 by adding color jittering. But for segmentation of two other lesions color jittering have no significant improvement on test dice, with an increase of 0.0014 and -0.0358 respectively in the 8th and 6th rows of Table \ref{table:result task2 of NPA} and the 5th and 4th rows of Tabel \ref{table:result task2 of NV}.

We investigate the effectiveness of color jittering (CJ) augmentation. For IRMAs segmentation, applying CJ can greatly improve the performance on test dice by 7.40\%, as can be found in the 5th and 6th rows of Table \ref{table:result task2 of IRMA}. The results demonstrate that CJ is effective to improve the generalization ability of the model. However, for the segmentation of NPAs and NV, we experimentally find that CJ shows no significant improvement on test dice.

\subsubsection{Metrics.}
% Because of evaluating the algorithm of segmentation by Dice and IoU in this challenge, we tried the two metrics to find best model. As can be seen from the row 1st and 2nd rows of Table \ref{table:result task2 of IRMA} and the 1st and 2nd rows of Table  \ref{table:result task2 of NV}, the test Dice only has 0.0006 and 0.0132 difference respectively. It also can be found in the leaderboard task1 of this challenge, 
% Dice's improvement is consistent with IoU's improvement.

Following the same protocol as DRAC challenge, we adopt Dice and IoU metrics to evaluate our model on the validation set and choose the best model. As can be observed in the first two rows of all the three tables, the models measured by the two metrics obtain very close segmentation performance for each lesion.

\subsubsection{Model ensemble.}
To further boost the performance on test samples, we ensemble several models for IRMAs and NPAs segmentation. As shown in Table \ref{table:result task2 of IRMA}, the ensembled model composed of model ID `1,2,4' (7th row) improves the performance by 0.85\% test dice, compared to the best UNet++ (1st row). In the 6th and 9th rows of Table \ref{table:result task2 of NPA}, we can observe that the ensembled model of model ID `4,5,6,7' obtains higher test dice than the best UNet by a 1.39\% improvement.

% For segmentation of IRMAs and NPAs, we ensemble several models to improve the performance on test dice. Comparing with best single model, the ensemble model improve 0.0085 and 0.0139 of test Dice respectively in the 1st and 8nd rows of Table \ref{table:result task2 of IRMA} and the 7th and 10th rows of Table \ref{table:result task2 of NPA}.

\subsubsection{Results on the leaderboard.}
Table \ref{table:leaderboard task1} shows the top-5 results of our method and other participants on the testing set of Task 1 in the DRAC challenge. Our ensembled UNet++ for IRMAs, ensembled UNet and UNet++ for NPAs, and single model UNet for NV, with customized learning rate schedules and color jittering augmentation ranked 4th on the leaderboard, approaching 0.5491 mDice and 0.4041 mIoU.

\section{Image Quality Assessment}

\subsection{Methodology}
Our proposed model takes an OCTA image $X\in \mathbbm{R}^{W\times H\times3}$ as the input and outputs the image quality prediction $\hat{Y}$ in an end-to-end manner. To alleviate the overfitting problem caused by limited samples, we first pre-train our model on the large-scale OCTA-25K-IQA-SEG dataset. Then, we fine-tune the model on DRAC dataset using the hybrid MixUp and CutMix strategy. We create an ensemble of three models, including Inception-V3, SE-ResNeXt, and Vision Transformer (ViT). In the following section, we will introduce the three models, hybrid MixUp and CutMix strategy, and joint loss function in detail.

\subsubsection{Backbone network.}
(1) Inception-V3 \cite{szegedy2016rethinking} aims to reduce the parameter count and computational cost by replacing larger convolutions with a sequence of smaller and asymmetric convolution layers. It also adopts auxiliary classifiers with batch normalization as aggressive regularization.
(2) SE-ResNeXt \cite{xie2017aggregated} is composed of SENet and ResNeXt, where SENet employs the Squeeze-and-Excitation block to recalibrates channel-wise feature responses adaptively, and ResNeXt aggregates a set of transformations with the same topology.
(3) ViT \cite{dosovitskiy2020image} receives a sequence of flattened image patches as input into a Transformer encoder,  which consists of alternating layers of multi-head attention and MLP blocks.

\subsubsection{Hybrid MixUp and CutMix strategy.}
We adopt the MixUp strategy \cite{zhang2017mixup} to mix training data, where each sample is interpolated with another sample randomly chosen from a mini-batch. Specifically, for each pair of OCTA images $(x_i, x_j)$ and their quality labels $(y_i, y_j)$, the mixed $(x', y')$ is computed by:
\begin{equation}
    \begin{aligned}
    & \lambda \sim \textrm{Beta}(\alpha_1, \alpha_1),\\
    & x'=\lambda x_i + (1-\lambda) x_j,\\
    & y'=\lambda y_i + (1-\lambda) y_j,
    \end{aligned}
\end{equation}
where the combination ratio $\lambda\in [0,1]$ is sampled from the beta distribution. The hyper-parameter $\alpha_1$ controls the strength of interpolation of each pairs, and we set $\alpha_1=0.4$ in this work.

We also utilize another augmentation strategy named CutMix \cite{yun2019cutmix}. It first cuts patches and then pastes them among training images, where the labels are also mixed proportionally to the area of the patches. Formally, let $(x_i, y_i)$ and $(x_j, y_j)$ denote two training samples. The combined $(x', y')$ is generated as:
\begin{equation}
    \begin{aligned}
    & \lambda \sim \textrm{Beta}(\alpha_2, \alpha_2),\\    
    & x'=\mathbf{M}\odot x_i+(\mathbf{1}-\mathbf{M})\odot x_j,\\
    & y'=\lambda y_i + (1-\lambda) y_j,
    \end{aligned}
\end{equation}
where $\mathbf{M}\in \{0,1\}^{W\times H}$ denotes a binary mask indicating where to drop out and fill in, $\mathbf{1}$ is a matrix filled with ones, and $\odot$ is element-wise multiplication. In our experiments, we set $\alpha_2$ to 1. 
The mask $\mathbf{M}$ can be determined by the bounding box coordinates $\mathbf{B}=(r_x, r_y, r_w, r_h)$, which are uniformly sampled according to:
\begin{equation}
    \begin{aligned}
        & r_x \sim \mathrm{Unif}(0,W),~ r_y \sim \mathrm{Unif}(0,H),\\
        & r_w=W\sqrt{1-\lambda},~ r_h=H\sqrt{1-\lambda}.
    \end{aligned}
\end{equation}
With the cropping region, the binary mask $\mathbf{M}\in \{0,1\}^{W\times H}$ is decided by filling with 0 within the bounding box $\mathbf{B}$, otherwise 1.

% In each training iteration, a CutMix-ed sample is generated by combining randomly selected two training samples in a mini-batch.

\subsubsection{Joint loss function.}
Different from the original design where they replaced the classification loss with the mix loss, we merge the standard cross-entropy classification loss $\mathcal{L}_{clf}$ with the mix loss $\mathcal{L}_{mix}$ to enhance the classification ability on both raw and mixed samples. The joint loss can be calculated as $\mathcal{L}=\mathcal{L}_{clf}+\mathcal{L}_{mix}$.
% \begin{equation}
%     \mathcal{L}=\mathcal{L}_{clf}+\mathcal{L}_{mix}.
% \end{equation}
Specifically, the classification loss is expressed by:
\begin{equation}
    \mathcal{L}_{clf}=-\frac{1}{N}\sum_{i=1}^N y_i^T\log \hat{y}_i,
\end{equation}
where $N$ denotes the total number of samples, $y_i$ is the one-hot vector of ground truth label, and $\hat{y}_i$ is the predicted probability of sample $x_i$. For the mix loss, we additionally utilize label smoothing \cite{szegedy2016rethinking} to reduce overfitting:
\begin{equation}
    \mathcal{L}_{mix}=-\frac{1}{N}\sum_{i=1}^N \tilde{y}_i^{'T}\log \hat{y}'_i,
\end{equation}
where $\hat{y}'_i$ is the predicted probability of mixed sample $x'_i$, and $\tilde{y}'_i=y_i'(1-\epsilon)+\epsilon/K$ represents the smoothed label, $K$ denotes the number of classes. In our work, the hyper-parameter $\epsilon$ is set to $0.1$.

\subsection{Dataset}
\subsubsection{DRAC dataset.}
The DRAC dataset for image quality assessment contains three different levels, i.e. Poor quality level (0), Good quality level (1), and Excellent quality level (2). Training set consists of 665 images and corresponding labels, where each grade includes 50/97/518, respectively. We use 5-fold cross validation. Testing set consists of 438 images and the labels are not available during the challenge. 
\subsubsection{OCTA-25K-IQA-SEG dataset.}
The OCTA-25K-IQA-SEG dataset \cite{wang2021deep} provides 14,042 $6\times 6 \mathrm{mm}^2$ superficial vascular layer OCTA images with annotations of image quality assessment (IQA) and foveal avascular zone (FAZ) segmentation. Similarly, each OCTA image is labeled into three IQA categories, including ungradable (0), gradable (1), or outstanding (2). In this work, we adopt the OCTA-25K-IQA-SEG dataset for network pre-training.

\subsection{Implementation details}

All images are resized to $224\times 224$ as the input to SE-ResNeXt, $384\times 384$ for ViT, and $512\times 512$ for other models. Data augmentation includes random cropping, flipping, and color jittering. 
The networks are optimized using the SGD algorithm and trained for 100 epochs. The initial learning rate is set to 1e-3, and we use cosine annealing learning rate schedule. 

\subsection{Experimental results}

\setlength{\tabcolsep}{4pt}
\begin{table}[t]
\begin{center}
\caption{The results of image quality assessment on the DRAC dataset. The val Kappa is the mean value of 5-fold cross validation.}
\label{table:val task2}
\resizebox{\linewidth}{!}{
\begin{tabular}{c|l|c|c|c|c}
\hline
% \noalign{\smallskip}
ID & Method & Pre-training & Mix & Val Kappa & Test Kappa\\
% \noalign{\smallskip}
\hline
% \noalign{\smallskip}
1 & Inception-V3 \cite{szegedy2016rethinking} & \multirow{6}{*}{ImageNet} & $\times$ & 0.8203 & 0.7817\\
2 & Inception-Res-V2 \cite{szegedy2017inception} & & $\times$ & 0.8222 & 0.6881\\
3 & EfficientNet-B6 \cite{tan2019efficientnet} & & $\times$ & 0.7866 & Unknown\\
4 & ResNeSt-50 \cite{zhang2022resnest} & & $\times$ & 0.8450 & 0.7444\\
5 & ViT-t \cite{dosovitskiy2020image} & & $\times$ & 0.8488 & 0.7253\\
6 & ViT-s \cite{dosovitskiy2020image} & & $\times$ & 0.8272 & 0.7298\\
\hline
7 & \multirow{2}{*}{SE-ResNeXt-101 \cite{hu2018squeeze}} & \multirow{6}{*}{OCTA-25K-IQA-SEG} & $\times$ & 0.8774 & 0.7580\\
8 & & & $\checkmark$ & 0.8560 & 0.7647\\
\cline{2-2}\cline{4-6}
9 & \multirow{2}{*}{Inception-v3 \cite{szegedy2016rethinking}} & & $\times$ & 0.8576 & 0.7358 \\
10 & & & $\checkmark$ & 0.8189 & 0.6989 \\
\cline{2-2}\cline{4-6}
11 & \multirow{2}{*}{ViT-s \cite{dosovitskiy2020image}} & & $\times$ & 0.8621 & Unknown\\
12 & & & $\checkmark$ & 0.8710 & Unknown\\
\hline
13 & Ensemble ID `1, 4, 5, 6' & - & - & - & 0.7835\\
14 & Ensemble ID `1, 12' & - & - & - & 0.7864\\
15 & Ensemble ID `1, 5, 6' & - & - & - & 0.7877\\
16 & Ensemble ID `1, 8, 12' & - & - & - & \textbf{0.7896}\\
\hline
\end{tabular}
}
\end{center}
\end{table}
\setlength{\tabcolsep}{1.4pt}

\subsubsection{Backbone network.}
First, we evaluate the effects of different backbone architectures. As can be seen from the 1st to 6th rows of Table \ref{table:val task2}, the ResNeSt-50 \cite{zhang2022resnest} and ViT-t \cite{dosovitskiy2020image} models achieve superior performance with more than 0.84 Kappa, and the Inception-V3 \cite{szegedy2016rethinking}, Inception-Res-V2 \cite{szegedy2017inception} and ViT-s \cite{dosovitskiy2020image} models reach over 0.82 Kappa on the 5-fold cross validation. Surprisingly, the Inception-V3 outperforms other models on the testing set, approaching 0.7817 on Kappa score. 

\subsubsection{Pre-training scheme.}
To alleviate the overfitting issue, we first pre-train our models on the large-scale OCTA-25K-IQA-SEG dataset \cite{wang2021deep}. From the validation results in the 7th, 9th and 11th rows of Table \ref{table:val task2}, we find the pre-trained model can offer better initialization weights. Compared to ImageNet pre-training, the Inception-V3 and ViT-s models achieve 3.73\% and 3.49\% improvements on val Kappa, respectively. The SE-ResNeXt-101 \cite{hu2018squeeze} reaches the best performance with 0.8774 val Kappa. 

\subsubsection{Hybrid MixUp and CutMix strategy.}
In addition, we investigate the impact of hybrid Mixup and CutMix strategy in the 7th to 12th rows of Table \ref{table:val task2}. It can be seen that this strategy successfully increases the test Kappa by 0.67\% for SE-ResNeXt-101, as well as the val Kappa by 0.89\% for ViT-s. However, we experimentally find that the strategy cannot benefit the performance of Inception-V3.

\subsubsection{Model ensemble.}
Finally, we ensemble several models to further boost the performance of image quality assessment. The final prediction of each OCTA image is obtained by averaging the predictions from individual models. We show the results of different model combinations in the 13th to 16th rows of Table \ref{table:val task2}. The ensembled model, which is composed of ID `1,8,12' models, surpasses all the other models, approaching 0.7896 Kappa on the testing set.

\subsubsection{Results on the leaderboard.}
Table \ref{table:leaderboard task2} presents the top-5 results of our method and other participants on the testing set of Task 2 in the DRAC challenge. Our final method, an ensemble of Inception-V3, SE-ResNeXt-101, and ViT-s, ranked 3rd on the leaderboard with 0.9083 AUC and 0.7896 Kappa.

\setlength{\tabcolsep}{4pt}
\begin{table}[t]
\begin{center}
\caption{The top-5 results on the leaderboard of image quality assessment.}
\label{table:leaderboard task2}
% \resizebox{\linewidth}{!}{
\begin{tabular}{llcc}
\hline\noalign{\smallskip}
Rank & Team & AUC & Kappa \\
\noalign{\smallskip}
\hline
\noalign{\smallskip}
1 & FAI & 0.8238 & 0.8090 \\
2 & KT\_Bio\_Health & 0.9085 & 0.8075 \\
3 & FDVTS\_DR (ours) & 0.9083 & 0.7896 \\
4 & LaTIM & 0.8758 & 0.7804 \\
5 & yxc & 0.8874 & 0.7775 \\
\hline
\end{tabular}
% }
\end{center}
\end{table}
\setlength{\tabcolsep}{1.4pt}

\section{Diabetic Retinopathy Grading}

\subsection{Methodology}
We adopt a Vision Transformer (ViT) model for DR grading, which receives an OCTA image and produces the grading score in an end-to-end manner. Following the same training procedure in Task 2, we first pre-train our model on EyePACS \cite{kaggle} and DDR \cite{li2019diagnostic} datasets, which provide a large number of color fundus images for DR grading. We then fine-tune the model on the DRAC dataset with the hybrid MixUp and CutMix strategy. The model is trained to minimize a joint objective function of a mix loss and a classification loss.

\subsection{Dataset}
\subsubsection{DRAC dataset.}
The DRAC dataset for diabetic retinopathy grading contains three grades, i.e. Normal (0), NPDR (1), PDR (2). Training set consists of 611 images and corresponding labels, where each grade includes 329/212/70, respectively. We use 5-fold cross validation. Testing set consists of 386 images and the labels are not available during the challenge. 

\subsubsection{Fundus image datasets.}
We employ two large-scale datasets, i.e. EyePACS dataset \cite{kaggle} and DDR dataset \cite{li2019diagnostic}, which offer color fundus images for DR grading. Each image is categorized into five grades, including No DR, Mild NPDR, Moderate NPDR, Severe NPDR, and PDR. We adopt training sets of the two datasets, composed of 35,126 and 6,260 images for network pre-training.

\subsection{Implementation details}
All images are resized to $384\times 384$ as the input for ViT, and $512\times 512$ for other models. Random cropping, flipping, rotation, and color jittering are applied on the training data. 
We train the networks for 100 epochs using the SGD algorithm. The initial learning rate is set to 1e-3, and decay by cosine annealing learning rate schedule. 

\subsection{Experimental results}

\subsubsection{Backbone network.} 
As can be seen in Table \ref{table:val task3}, we compare the performance of different networks, including Inception-V3 \cite{szegedy2016rethinking}, ResNeSt-50 \cite{zhang2022resnest}, ViT-t \cite{dosovitskiy2020image}, and ViT-s \cite{dosovitskiy2020image}. Among the four networks, ViT-s shows the best performance with 0.8486 val Kappa and 0.8365 test Kappa. The results demonstrate the ability of ViT to capture long-range spatial correlations within each image. Therefore, we choose ViT-s as our backbone network for DR grading.

\setlength{\tabcolsep}{4pt}
\begin{table}[t]
\begin{center}
\caption{The results of diabetic retinopathy grading on the DRAC dataset. The val Kappa is the mean value of 5-fold cross validation.}
\label{table:val task3}
% \resizebox{\linewidth}{!}{
\begin{tabular}{c|l|c|c|c|c}
\hline
% \noalign{\smallskip}
ID & Method & Pre-training & Mix & Val Kappa & Test Kappa\\
% \noalign{\smallskip}
\hline
% \noalign{\smallskip}
1 & Inception-V3 \cite{szegedy2016rethinking} & \multirow{4}{*}{ImageNet} & $\times$ & 0.8366 & 0.7445\\
2 & ResNeSt-50 \cite{zhang2022resnest} & & $\times$ & 0.8214 & Unknown\\
3 & ViT-t \cite{dosovitskiy2020image} & & $\times$ & 0.8433 & 0.7860\\
4 & ViT-s \cite{dosovitskiy2020image} & & $\times$ & 0.8486 & 0.8365\\
\hline
5 & \multirow{4}{*}{ViT-s \cite{dosovitskiy2020image}} & \multirow{4}{*}{EyePACS \& DDR} & 0.0 & 0.8599 & 0.8539 \\
6 & & & 0.1 & 0.8656 & 0.8639 \\
7 & & & 0.5 & 0.8624 & \textbf{0.8693}\\
8 & & & 1.0 & 0.8625 & 0.8427\\
\hline
13 & TTA (t=2) & \multirow{3}{*}{EyePACS \& DDR} & \multirow{3}{*}{0.5} & - & 0.8636\\
14 & TTA (t=3) & & & - & 0.8664\\
15 & TTA (t=4) & & & - & 0.8592\\
\hline
\end{tabular}
% }
\end{center}
\end{table}
\setlength{\tabcolsep}{1.4pt}

\subsubsection{Pre-training scheme.}

As the number of OCTA images on DRAC dataset is relatively limited, we additionally adopt EyePACS \cite{kaggle} and DDR \cite{li2019diagnostic} datasets to pre-train the networks, which provide a large number of color fundus images for DR drading. 
Compared to the ImageNet pre-trained ViT-s (4th row), the EyePACS \& DDR pre-trained ViT-s (5th row) significantly improves the performance by 1.13\% val Kappa and 1.74\% test Kappa. The results illustrate that the representation of color fundus images can serve as a useful substrate for DR grading on OCTA images.

\subsubsection{Hybrid MixUp and CutMix strategy.}
Moreover, the hybrid MixUp and CutMix strategy also greatly boosts the DR grading performance, as shown in the 6th to 8th rows of Table \ref{table:val task3}. We set different values of mix probability, which controls the probability of a sample being mixed. When the probability is set to $0.1$ (6th row), the ViT-s achieves 0.57\% and 1\% improvements on val and test Kappa, respectively. With the increase of probability, the test Kappa continues to improve and reaches 0.8693 when the probability is $0.5$. However, if the value of probability is set too large (e.g. $1.0$), the training phase would be difficult to converge, which leads to inferior grading performance (i.e. 0.8427 test Kappa). According to the results, we choose the best-performing ViT-s with the mix probability $0.5$ as our final model.

\subsubsection{Test time augmentation.}
Aiming to further boost the generalization ability of our model on the testing set, we employ a test time augmentation (TTA) operation. For each image $x$, we apply $t$-time augmentations $\mathcal{A}$ and obtain $\{\mathcal{A}_1(x), \mathcal{A}_2(x), ..., \mathcal{A}_t(x)\}$. Then, we average the model’s predictions on these samples. 
The last three rows of Table \ref{table:val task3} show the results of TTA with different $t$ (i.e. $t=2,3,4$). However, it is observed that our augmentations do not further promote the grading performance. It is probably because our strong augmentation is beneficial to the training phase but may be harmful in testing phase.

\setlength{\tabcolsep}{4pt}
\begin{table}[t]
\begin{center}
\caption{The top-5 results on the leaderboard of diabetic retinopathy grading.}
\label{table:leaderboard task3}
% \resizebox{\linewidth}{!}{
\begin{tabular}{llcc}
\hline\noalign{\smallskip}
Rank & Team & AUC & Kappa \\
\noalign{\smallskip}
\hline
\noalign{\smallskip}
1 & FAI & 0.9147 & 0.8910 \\
2 & KT\_Bio\_Health & 0.9257 & 0.8902 \\
3 & LaTIM & 0.8938 & 0.8761 \\
4 & BUAA\_Aladinet & 0.9188 & 0.8721 \\
5 & FDVTS\_DR (ours) & 0.9239 & 0.8636 \\
\hline
\end{tabular}
% }
\end{center}
\end{table}
\setlength{\tabcolsep}{1.4pt}

\subsubsection{Results on the leaderboard.}
Table \ref{table:leaderboard task3} presents the top-5 results of our method and other participants on the testing set of Task 3 in the DRAC challenge. Our EyePACS \& DDR pre-trained ViT-s model with Mix and TTA strategies ranked 5th on the leaderboard, approaching 0.9239 AUC and 0.8636 Kappa.

\section{Conclusion}
In this work, we presented our solutions of three tasks in the Diabetic Retinopathy Analysis Challenge. 
In task 1, we adopted UNet and UNet++ networks with pre-trained encoders for DR lesion segmentation, which benefited from a series of strong data augmentation.
In task 2, we created an ensemble of various networks for image quality assessment. Both the pre-training scheme and hybrid MixUp and CutMix strategy helped to boost the performance. 
In task 3, we developed a ViT model for DR grading. We found the model pre-trained on color fundus images serves as a useful substrate for OCTA images.
The experimental results demonstrated the effectiveness of our proposed methods. On the three leaderboards, our methods ranked 4th, 3rd, and 5th, respectively.

\bibliographystyle{splncs04}
\bibliography{mybibliography}
% 
% \begin{thebibliography}{8}
% \bibitem{ref_article1}
% Author, F.: Article title. Journal \textbf{2}(5), 99--110 (2016)

% \bibitem{ref_lncs1}
% Author, F., Author, S.: Title of a proceedings paper. In: Editor,
% F., Editor, S. (eds.) CONFERENCE 2016, LNCS, vol. 9999, pp. 1--13.
% Springer, Heidelberg (2016). \doi{10.10007/1234567890}

% \bibitem{ref_book1}
% Author, F., Author, S., Author, T.: Book title. 2nd edn. Publisher,
% Location (1999)

% \bibitem{ref_proc1}
% Author, A.-B.: Contribution title. In: 9th International Proceedings
% on Proceedings, pp. 1--2. Publisher, Location (2010)

% \bibitem{ref_url1}
% LNCS Homepage, \url{http://www.springer.com/lncs}. Last accessed 4
% Oct 2017
% \end{thebibliography}
\end{document}